\documentclass[showpacs,nofootinbib]{revtex4}

\usepackage{amsfonts}
\usepackage{amsbsy} 
\usepackage{epsfig}
\usepackage{latexsym}
\usepackage{dbnsymb}
\input amssym.def
\input amssym.tex

\def\half{{1\over 2}}
\def\ben{\begin{equation}}
\def\een{\end{equation}}
\def\bea{\begin{eqnarray}}
\def\nn{\nonumber}
\def\eea{\end{eqnarray}}
\newcommand{\sect}[1]{\setcounter{equation}{0}\section{#1}}

\advance \topmargin by -\headheight
\advance \topmargin by -\headsep
\evensidemargin \oddsidemargin
\advance\hoffset by -3mm  % A4 is narrower.
\advance\voffset by  8mm  % A4 is taller.
\def\const{\rm constant}

\def\ie{{\it i.e. }}

\def\sgn{{\rm sgn }}

\def\bR{\Bbb R}

\def\b1{e^0}

\begin{document}
\title{Classical general relativity as BF-Plebanski theory with linear constraints}

\begin{flushright}DAMTP-2010-34, AEI-2010-084\end{flushright}

\author{{\bf Steffen Gielen}\footnote{sg452@cam.ac.uk}}
\affiliation{DAMTP, Centre for Mathematical Sciences, Cambridge University, Wilberforce Road, Cambridge CB3 0WA, U.K., EU} 
\author{{\bf Daniele Oriti}\footnote{daniele.oriti@aei.mpg.de}}
\affiliation{Max Planck Institute for Gravitational Physics (Albert Einstein Institute) \\ Am M\"uhlenberg 1, D-14476 Golm, Germany, EU}
\begin{abstract}
We investigate a formulation of continuum 4d gravity in terms of a constrained BF theory, in the spirit of the Plebanski formulation, but involving only linear constraints, of the type used recently in the spin foam approach to quantum gravity. We identify both the continuum version of the linear simplicity constraints used in the quantum discrete context and a linear version of the quadratic volume constraints that are necessary to complete the reduction from the topological theory to gravity. We illustrate and discuss also the discrete counterpart of the same continuum linear constraints. Moreover, we show under which additional conditions the discrete volume constraints follow from the simplicity constraints, thus playing the role of secondary constraints. 
\end{abstract}


\maketitle
\sect{Introduction}
The equations of general relativity can be derived from several different action principles \cite{peldan}, leading to equivalent classical theories (in the case of pure gravity, at least). Among them we can mention, in addition to the Einstein-Hilbert action \cite{hilbert}, the Palatini first order formulation and its modification proposed by Holst \cite{holst}. This last one is of special interest because it is the classical, covariant starting point for the canonical quantisation leading to Loop Quantum Gravity \cite{carlo, thomas}. Even if not the only possible useful one \cite{leeartem}, a particularly popular action in covariant approaches to quantising gravity \cite{libro}, like the spin foam \cite{SF} and group field theory approach \cite{iogft}, is the formulation as a constrained BF (or Plebanski \cite{plebanski, kirill}) theory. Here one starts from topological BF theory \cite{horowitzBF} in 4 spacetime dimensions, and adds suitable constraints on the two-form $B$ variables of the theory such that, on solutions of these constraints, the action reduces to the Palatini or Holst action for general relativity. We will summarise the idea behind this formulation in the following. In the original Plebanski formulation the constraints on the $B$ variables are quadratic, and so are the discrete constraints that are then implemented in the spin foam models based on a simplicial discretisation. On the other hand, the most recent developments in the spin foam and group field theory approach to quantum gravity are based on a linear set of discrete constraints, which can be shown to be slightly stronger, in the restrictions they impose on the original BF configurations, than some of the original discrete quadratic constraints. Once more, we will detail this construction in the following. 

\

In this note we investigate whether a formulation in terms of linear constraints is also possible in the classical continuum theory, and what it implies. We will see that the replacement of diagonal and cross-simplicity constraints with linear constraints  at the continuum level is relatively straightforward, after one has introduced new variables $n_A$ forming a basis of three-forms at each point. One then needs additional constraints corresponding to the volume constraints. We will see that one can linearise these constraints too. We then give a discrete version of these linear volume constraints, which bears a striking resemblance to the so-called ``edge simplicity" constraints of \cite{biancajimmy}. We note that only certain linear combinations of the volume constraints one would naively write down are necessary to constrain the bivectors $\Sigma^{AB}(\triangle)$ sufficiently.
Similarly to the quadratic case, we will also see that when linear diagonal and off-diagonal constraints hold everywhere in a 4-simplex, and one also imposes closure constraints on both bivectors $\Sigma^{AB}(\triangle)$ and normals $n_A(\tetrahedron)$ referring to the tetrahedra of the 4-simplex, the sufficient set of linear combinations of the linear volume constraints follows. This additional four-dimensional closure constraint on the normal vectors has, to the best of our knowledge, not been considered or implemented as an additional condition in the spin foam literature yet, although it does appear in some first order formulation of Regge calculus \cite{caselle}, and it plays also a role in the discrete analysis of \cite{FreidelConradysemiclassical}.

\sect{Gravity as constrained BF theory: continuum and discrete results}
\label{intro}
Let us briefly review what is known at the classical continuum and discrete level, concerning the Plebanski formulation of classical gravity. We limit our considerations to the covariant, Lagrangian context, and to a very small subset of the available results, those which have been already of direct relevance for quantum gravity model building, especially in the spin foam context. For recent results in the canonical Hamiltonian setting, both continuum and discrete, see \cite{henneaux, biancajimmy, zapata, sergeikirill}.

\

Consider first the Einstein-Hilbert-Palatini-Holst \cite{hilbert,holst} Lagrangian (without cosmological constant),\footnote{We follow the usual conventions: Capital Latin indices are internal indices, small Latin indices are coordinate indices. For an abstract collection of indices, we use small Greek letters.}
\ben
S = \frac{1}{8\pi G}\int_{\mathcal{R}\times\mathbb{R}}\left(\half\,\epsilon_{ABCD}\,E^A\wedge E^B\wedge R^{CD}[\omega]+\frac{1}{\gamma}\,E^A\wedge E^B\wedge R_{AB}[\omega]\right),
\label{einsthilb}
\een
where spacetime is assumed to be of the form $\mathcal{R}\times\mathbb{R}$ so that a (3+1) splitting can be performed, $\omega^{AB}$ is a $G$-connection one-form (the gauge group $G$ is $SO(3,1)$ or $SO(4)$, or an appropriate cover), $R^{AB}$ its curvature, and $E^A$ is an $\bR^4$-valued one-form representing an orthonormal frame. The term involving $\gamma$, known as the Holst term, is not relevant classically; its variation vanishes if torsion is zero (but see \cite{mercuri}). It is, however, of fundamental importance in loop quantum gravity (LQG), and more generally for any canonical formulation of gravity, as it modifies the symplectic structure of the theory.

If one introduces a $\frak{g}$-valued two-form,
\ben
B^{AB}=\frac{1}{8\pi G}\left(\half{\epsilon^{AB}}_{CD}\,E^C\wedge E^D+\frac{1}{\gamma}E^A\wedge E^B\right),
\label{bfield}
\een
then the action (\ref{einsthilb}) becomes
\ben
S = \int B^{AB}\wedge R_{AB}[\omega]+\lambda^{\alpha}C_{\alpha}[B],
\een
\ie it takes the form of a topological BF theory with additional constraints $C_{\alpha}$ which enforce that $B^{AB}$ is indeed of the form (\ref{bfield}), and that are enforced by means of Lagrange multipliers $\lambda_\alpha$. 

As said, BF theory without constraints is topological. Its equations of motion imply that $\omega^{AB}$ is flat and the covariant exterior derivative of $B^{AB}$ vanishes. Having no local degrees of freedom, the quantisation of such a theory is therefore rather simple and quite well understood. Inspired by this classical formulation, the main issue when trying to construct a quantum theory related to quantum gravity, in 4 dimensions\footnote{Notice that, in 3 spacetime dimensions, gravity in 1st order form coincides, classically, with 3d BF theory, and that in higher dimensions a similar formulation of gravity as constrained BF theory can be given \cite{FKhigherBF}.}, is then the correct implementation of appropriate constraints that lead to (\ref{bfield}) for some set of one-forms $E^A$, either at the level of quantum states or in a path integral formulation. Indeed, the bulk of the work in the spin foam approach \cite{SF, eprlong, eprl, consistent, freidelkrasnov} (as well as in the group field theory formalism \cite{iogft, danieleGFtsimpl, danieleAristide}), in recent years, has been devoted to this task.
These constraints are also the subject of this note.

To simplify the following calculations, we introduce another two-form field $\Sigma^{AB}$,
\ben
\Sigma^{AB}\equiv\frac{1}{1-s\gamma^2}\left(B^{AB}-\frac{\gamma}{2}{\epsilon^{AB}}_{CD}B^{CD}\right) \label{redefB},
\een
where $s$ is the spacetime signature, $s=-1$ for $G=SO(3,1)$ and $s=+1$ for $G=SO(4)$ (and we assume $\gamma^2 \neq s$)\footnote{For uniformity of the discussion, we shall in the following talk about ``time'' and use the label $0$ even when the gauge group is $SO(4)$ and the signature Riemannian.}. This is a linear redefinition which simplifies the constraint (\ref{bfield}),
\ben
\Sigma^{AB} = \frac{1}{8\pi\gamma G}\,E^A\wedge E^B,
\label{sigma}
\een
but leads to more terms in the action. The translation of all calculations from one set of variables to the other is usually straightforward.

The traditional way to enforce the restriction (\ref{sigma}), the one matching the original classical Plebanski formulation of gravity, was to add quadratic {\it simplicity constraints}\footnote{``Simplicity'' because a two-form that can be written as a wedge product of one-forms is called simple.} to the action \cite{reisenclass,freidelpleb},
\ben
\epsilon_{ABCD}\Sigma^{AB}_{ab}\Sigma^{CD}_{cd}=V\epsilon_{abcd},
\label{quadconst}
\een
where $V$ can be expressed in terms of $\Sigma^{AB}$ by contracting (\ref{quadconst}) with $\epsilon^{abcd}$, to give: $V=\frac{s}{24} \epsilon^{abcd}\epsilon_{ABCD}\Sigma^{AB}_{ab}\Sigma^{CD}_{cd}$. This is itself a reformulation of the original Plebanski constraint, which would read:
\ben
\epsilon^{abcd}\Sigma^{AB}_{ab}\Sigma^{CD}_{cd}=V\epsilon^{ABCD},
\label{quadconstB}
\een
and is equivalent to the first under assumption that $V\neq 0$ everywhere. 
The version (\ref{quadconst}) has the advantage of permitting a much simpler discretisation and thus a more straightforward implementation within the spin foam formalism. 
Under the same assumption $V\neq 0$, there are the following four classes of solutions to (\ref{quadconst}):
\ben
\mbox{either }\;\Sigma^{AB}=\pm e^A\wedge e^B\quad\mbox{or }\;\Sigma^{AB}=\pm\half{\epsilon^{AB}}_{CD}e^C\wedge e^D
\label{ambiguity}
\een
for some set of one-forms $e^A$ (the factor $8\pi\gamma G$ can obviously be introduced by rescaling). One would like to select only the first class of solutions $\Sigma^{AB}=+e^A\wedge e^B$, which, when substituted in the BF action, gives the Holst action (\ref{einsthilb}). Classically, this is not a severe problem. As shown in \cite{reisenclass}, non-degenerate initial data of a solution of the form $\Sigma^{AB}=+e^A\wedge e^B$ generically remain within the same branch of solutions. The situation in the quantum theory, where one necessarily has contributions from all branches, is less clear. 

More troublesome, if $V=0$, the field $\Sigma^{AB}$ does not permit a straightforward geometric interpretation at all. Since in the region of the phase space where $V=0$, the theory is less constrained, and hence has more degrees of freedom, these non-geometric configurations should be expected to be dominating in a path integral \cite{reisenclass}, unless measure factors are such that this is avoided. 

\

Spin foam models are usually defined in a piecewise flat context, and spin foam amplitudes are defined for given simplicial complexes \cite{SF}. Therefore one is interested in identifying a discrete version of the above constraints that could be imposed at the level of each complex. The version  (\ref{quadconstB}) of the simplicity constraints admits only a rather involved discrete counterpart \cite{freidelpleb} and, upon quantisation, leads to the Reisenberger model \cite{reisenberger,freidelpleb}, which has so far received only limited attention. 

The discrete analogue of the constraints (\ref{quadconst}) led instead \cite{reisenclass,freidelpleb} to the construction of the Barrett-Crane model \cite{barrettcrane}, in the case in which the Immirzi parameter is excluded from the original action ($\gamma\rightarrow\infty$). The construction is initially limited to a single 4-simplex, the convex hull of 5 points in $\mathbb{R}^4$ ($\mathbb{R}^{1,3}$, in the Lorentzian case) with the topology of a 4-ball, whose boundary is triangulated by the 5 tetrahedra identified by the 5 independent subsets of 4 such points, while subsets of 3 points  identify the four triangles belonging to each of these 5 tetrahedra, each of the triangles being shared by a pair of tetrahedra.
One then associates a Lie algebra element (bivector) $\Sigma^{AB}_\triangle\in\mathfrak{so}(4)\simeq \wedge^2\mathbb{R}^4$ (similarly in the Lorentzian case) to each triangle $\triangle$ in a given triangulation by integrating the two-form $\Sigma^{AB}$ over $\triangle$. The task is then to constrain appropriately these Lie algebra variables (or their quantum counterpart) following the continuum treatment.

It is useful to split the set of continuum equations (\ref{quadconst}) into two sets. Out of the 21 equations (\ref{quadconst}), one first identifies and imposes those 18 which have zero on the right-hand side (the ``diagonal'' and ``off-diagonal'' simplicity constraints), 
\ben
\epsilon_{ABCD}\Sigma^{AB}_{ab}\Sigma^{CD}_{ab}=\epsilon_{ABCD}\Sigma^{AB}_{ab}\Sigma^{CD}_{ac}=0 \quad \forall \; a,b,c.
\label{diagonal}
\een

This corresponds to the case if one or two of the indices of the two fields $\Sigma$ coincide. At the discrete level, this translates into two triangles on which the same fields are discretised which either coincide or at least share a single edge, and thus belong to the same tetrahedron. Thus all bivectors $\Sigma^{AB}_\triangle$ are required to satisfy 
$$\epsilon_{ABCD} \Sigma_\triangle^{AB}\Sigma_\triangle^{CD}=0 \quad \text{(diagonal simplicity constraint)}$$ and $$\epsilon_{ABCD}\,\Sigma_\triangle^{AB}\,\Sigma_{\triangle'}^{CD}\,=\,0 \quad \text{ for all} \quad \triangle,\triangle' \quad \text{sharing an edge (cross-simplicity constraint)}. $$ These two sets of equations can be imposed at the level of each tetrahedron in the 4-simplex.

The remaining three equations (the ``volume'' constraints) are equivalent to the requirement that:
\ben
\epsilon_{ABCD}\Sigma^{AB}_{01}\Sigma^{CD}_{23}=\epsilon_{ABCD}\Sigma^{AB}_{02}\Sigma^{CD}_{13}=\epsilon_{ABCD}\Sigma^{AB}_{01}\Sigma^{CD}_{23}\,\propto\, V(\Sigma),
\label{quadvol}
\een
and can be imposed at the discrete level as the requirement that, for each 4-simplex:
\ben
\epsilon_{ABCD}\,\Sigma_\triangle^{AB}\,\Sigma_{\triangle'}^{CD}\,=\,V \quad \text{ for all} \quad \triangle,\triangle' \quad \textbf{not} \quad \text{sharing an edge (volume constraints)}
\een
where $V$ is defined by the above equation, and is interpreted, on the solutions of the constraints, as the volume of the 4-simplex.

An additional condition on the bivectors is usually considered, namely the ``closure'' constraint, which states that the sum of four bivectors corresponding to the faces of one tetrahedron is zero:
\ben
\sum_{\triangle\subset\tetrahedron} \Sigma_{\triangle}^{AB}\,=\,0\; . \label{closure}
\een
This constraint can be understood in two ways. 
One can either view it as the condition that the triangles described by the variables $\Sigma^{AB}_{\triangle}$ close to form a tetrahedron \cite{quanttetra}, or as a consequence of the equations of motion. In a topologically trivial region such as the interior of a tetrahedron, a flat connection can be set to zero by a gauge transformation. Then using Stokes' theorem, the integral over the equation $d\Sigma^{AB}=0$ can be written as $\int_{\tetrahedron}\Sigma^{AB}=0$, which is the closure constraint. The canonical counterpart of this condition is then the Gauss constraint, which generates local gauge (rotation) transformations and is to be imposed on the quantum states of the theory.

The same picture appears in three spacetime dimensions, where there are no simplicity constraints and one directly deals with a $\mathfrak{su}(2)$ connection one-form $\omega^{A}$ and an $\mathfrak{su}(2)$-valued one-form $e^A$. Here the equation $de^A=0$ is integrated over a (spacetime) triangle to give a closure constraint. The vectors (using $\mathfrak{su}(2)\simeq \mathbb{R}^3$) associated to the edges of the triangle add up to zero, and thus have a consistent geometric interpretation as edge vectors in $\mathbb{R}^3$. In this sense, an $n$-form with vanishing exterior derivative and appropriate internal indices can be given a geometric interpretation as describing $n$-simplices closing up to form an $(n+1)$-simplex. We shall encounter another instance of this statement later on.

The closure constraint, being linear in the $\Sigma$'s and local in each tetrahedron, is obviously easier to impose at the discrete level, and in the quantum theory, than the volume constraints. Thus it is a useful fact that it can indeed be imposed instead of them. More precisely, it can be shown  \cite{consistent} that the volume constraints in each 4-simplex are implied if one has enforced the diagonal and cross-diagonal simplicity constraints, {\it plus the closure conditions} everywhere, \ie in all the tetrahedra of the 4-simplex (in general, \ie for non-degenerate 4-simplices, involving tetrahedra belonging to different \lq\lq time slices\rq\rq). From a canonical perspective, this observation is usually phrased as an interpretation of the volume constraints as ``secondary constraints" required to guarantee conservation of the other constraints (including the Gauss (closure) constraint) under time evolution. 

After a period of investigations, several potentially worrying issues have been put forward regarding the Barrett-Crane model \cite{BCdegenerate, BCcoupling, alescirovelli} (for a more recent analysis of the geometry of the Barrett-Crane model, see \cite{danieleAristide}), and have given impetus to the development of alternative spin foam models \cite{eprlong,freidelkrasnov, eprl}. These models are known to have nice semiclassical properties \cite{nottingham}, and, importantly, generalise the spin foam setting to include the Immirzi parameter at the quantum level (for an early attempt, see \cite{danieleeteraImmirzi}), and thanks to this allow for a more direct contact with the canonical loop quantum gravity. Their study is still somewhat preliminary, but the above properties make them promising candidates for a quantum theory related to gravity. One of the central features of the new models is the replacement of the quadratic simplicity constraints (\ref{diagonal}) by linear constraints of the form
\ben
n_A(\tetrahedron)\Sigma^{AB}(\triangle)=0 \quad \forall\triangle\subset\tetrahedron\;,
\label{linear}
\een
where $n_A$ is the normal associated to the tetrahedron $\tetrahedron$ and $\triangle$ is any of the faces of $\tetrahedron$. 

It can be shown that these are lightly stronger than the discrete diagonal and off-diagonal quadratic simplicity constraints, and remove some of the discrete ambiguity in the solution for $\Sigma^{AB}$: out of the classes of solutions (\ref{ambiguity}), one can restrict to (the discrete version of) $\Sigma^{AB}=\pm e^A\wedge e^B$ only. For a geometric analysis of these conditions in the discrete setting, see \cite{eprlong, freidelkrasnov, FreidelConradysemiclassical, valentin}, and for a proof that the same discrete conditions can also lead to the Barrett-Crane model, see \cite{danieleAristide}.

\sect{Linear Constraints for BF-Plebanski Theory}
\label{constr}
The purpose of this note, as anticipated, is to investigate whether a formulation in terms of linear constraints is also possible in the classical continuum theory, and what it implies.

\

Let us work backwards, at first. Assume that the two-form field $\Sigma^{AB}$ is of the form $\Sigma^{AB}=e^A\wedge e^B$, and that the ``frame field'' $e^A$ is non-degenerate, \ie that the matrix $(e^A_a)$ is invertible. It follows that
\ben
e^c_A \Sigma^{AB}_{ab}=\delta_a^c e_b^B - \delta_b^c e_a^B.
\label{start}
\een
In order to make a connection to the discrete setting it is more convenient to work with exterior powers of the cotangent bundle only ($n$-forms can be integrated over $n$-dimensional submanifolds). Hence we multiply (\ref{start}) by $\epsilon_{cdef}$ and insert the relation $\epsilon_{cdef}e^c_A=\left(\det e^a_A\right)\epsilon_{ADEF}e^D_d e^E_e e^F_f$, which is true for invertible matrices, obtaining
\ben
\epsilon_{ADEF}e^D_d e^E_e e^F_f \Sigma^{AB}_{ab} = (\det e_a^A) \left(\epsilon_{adef}e_b^B-\epsilon_{bdef}e^B_a\right).
\label{linconst}
\een
One can define the three-form $n_{Adef}\equiv n_{A[def]}\equiv \epsilon_{ADEF}e^D_{d}e^E_{e}e^F_{f}$, so that (\ref{linconst}) take the form
\ben
n_{Adef} \Sigma^{AB}_{ab} = (\det e_a^A) \left(\epsilon_{adef}e_b^B-\epsilon_{bdef}e^B_a\right).
\label{betterform}
\een
$n_{Adef}$ can be interpreted as a 3d volume form for the submanifold parametrised by $(x^d,x^e,x^f)$ embedded in 4d spacetime, whose internal index gives the normal to this submanifold. If $e^A$ are a basis of one-forms at each spacetime point, then $n_A$ are a basis of three-forms at each spacetime point, and so one can choose to work either with one or the other set of variables. Clearly $e^A$ can be reconstructed from $n_A$:
\ben
\frac{1}{6}\epsilon^{cdef}n_{Adef}=s(\det e_a^A)e^c_A=s\sqrt[3]{\det\left(\frac{1}{6}\epsilon^{bdef}n_{Bdef}\right)}e^c_A.
\label{reconsta}
\een
This means that the set of variables $n_A(x)$ define a co-tetrad frame at any point of the spacetime manifold (for the discrete analogue of the above, see \cite{FreidelConradysemiclassical}).

\subsection{Linearised Diagonal and Off-Diagonal Constraints}

So far we have just rewritten the equation we want to obtain for $\Sigma^{AB}$. Let us now consider the implications of imposing (\ref{betterform}) as constraints, where we restrict to those with zero right-hand side, \ie those for which $\{a,b\}\subset\{d,e,f\}$. These are half of the equations (\ref{betterform}). This will identify the continuum analogue of the linear simplicity constraints.

{\bf Claim 1.} For a basis $n_{A}$ of three-forms, the general solution to
\ben
n_{Adef}\Sigma^{AB}_{ab}=0\quad\forall\{a,b\}\subset\{d,e,f\}
\label{zerorighthandside}
\een
is
\ben
\Sigma^{AB}_{ab}=G_{ab} e^{[A}_{a} e^{B]}_{b},
\label{solforb}
\een
where $e^A_a$ is defined in terms of $n_{Adef}$ as in (\ref{reconsta}), and assumed to be non-degenerate, and $G_{ab}=G_{ba}$ and $G_{aa}=0$. Obviously, as the variables $\Sigma$ and the tetrad field $e^A$, the ``coefficients'' $G_{ab}$ are spacetime dependent.

{\bf Proof.} First note that we can rewrite (\ref{zerorighthandside}) as
\ben
\epsilon_{ADEF}e^D_{d}e^E_{e}e^F_{f} \Sigma^{AB}_{ab} = 0
\label{zero2}
\een
with $e^A_a$ defined by (\ref{reconsta}). Then $e^A_a$ by assumption defines a basis in the cotangent space, so that
\ben
\Sigma^{AB}_{ab}=G^{gh}_{ab}e^A_{g}e^B_{h}
\een
for some coefficients $G^{gh}_{ab}$ with $G^{gh}_{ab}\equiv G^{[gh]}_{[ab]}$. Substituting this into (\ref{zero2}), we get
\ben
0\stackrel{!}{=}\epsilon_{ADEF}e^A_{g} e^D_{d}e^E_{e}e^F_{f} G^{gh}_{ab}e^B_{h} = \epsilon_{gdef}\det(e^A_{a})e^B_{h}G^{gh}_{ab},
\een
and since $\det (e^A_{a})\neq 0$ and $e^B_{h}$ form a basis of (the internal) $\mathbb{R}^4$, this implies that
\ben
\epsilon_{gdef}G^{gh}_{ab} = 0\quad\forall\{a,b\}\subset\{d,e,f\}.
\een
It follows that $G^{gh}_{ab}=0$ unless $\{g,h\}=\{a,b\}$ and so $G^{gh}_{ab}\equiv\delta^{[g}_{a}\delta^{h]}_{b}G_{ab}$.
\begin{flushright}
$\Box$
\end{flushright}

By a linear redefinition $e^A_a = \lambda_a \tilde{e}^A_a$ one might try to set some of the $G_{ab}$ to a given value (usually $\pm 1$, but one might prefer $\pm\frac{1}{8\pi\gamma G}$), but it is clear that one needs two conditions on the $G_{ab}$ for this to be possible.

In the discrete context, one sets $n_A(\tetrahedron)=(1,0,0,0)$ for each $\tetrahedron$ by a gauge transformation.\footnote{This presumably involves an implicit assumption, namely that there is a non-degenerate normal to each tetrahedron, as well.} One could use some of the gauge freedom here to restrict the form of $n_A$: This amounts to finding a convenient parametrisation for the coset space $GL(4)/SO(3,1)$. Let us make the (usual) assumption that the normal to hypersurfaces $\{t=\const\}$ is indeed timelike. Then one can use the boost part of $SO(3,1)$ to set $n_{A123}=(C,0,0,0)$. The remaining $SO(3)$ subgroup can then be used to make the ($3\times 3$) matrix $n_{I0de}$, where $I\in\{1,2,3\}$, upper diagonal, so that one has the form
\ben
n_{Adef}\sim\begin{pmatrix}{ * & * & * & * \cr 0 & * & * & * \cr 0 & 0 & * & * \cr 0 & 0 & 0 & * }\end{pmatrix}.
\een
Clearly, when this form of $n_{Adef}$ is assumed, integrating the three-form $n_A$ over a region where $t$ is constant gives a vector in $\mathbb{R}^4$ that only has a time component. Its magnitude specifies the three-dimensional volume of such a region.

\subsection{Linearised Volume Constraints}
As in the quadratic case, further constraints, in addition to the linear simplicity constraints (\ref{zerorighthandside}), are needed to complete the identification $\Sigma^{AB}=\pm E^A \wedge E^B$.

First of all, one can show the following.

{\bf Claim 2.} Under the assumption that all $G_{ab}$ are non-zero, the necessary and sufficient conditions for the existence of a linear redefinition $E^A_a=\lambda_a e^A_a$, such that either $\Sigma^{AB}=c E^A\wedge E^B$ or $\Sigma^{AB}=-c E^A\wedge E^B$, where $c$ is a given positive number, are
\ben
G_{12}G_{03}=G_{01}G_{23}=G_{13}G_{02}(\neq 0).
\label{gcondition}
\een

{\bf Proof.} Set $c=1$; the extension to arbitrary $c$ amounts to a further rescaling by $\sqrt{c}$. Then the required redefinition is possible if and only if there exist $\lambda_0,\ldots,\lambda_3$, such that either $G_{ab}=\lambda_a\lambda_b$ for all $a\neq b$ or $G_{ab}=-\lambda_a\lambda_b$ for all $a\neq b$. Clearly (\ref{gcondition}) are necessary. They are also sufficient: Take
\ben
\lambda_1=\sqrt{\left|\frac{G_{12}G_{13}}{G_{23}}\right|},\quad\lambda_2=\sgn\left(\frac{G_{12}G_{13}}{G_{23}}\right)\frac{G_{12}}{\lambda_1},\quad\lambda_3=\sgn\left(\frac{G_{12}G_{13}}{G_{23}}\right)\frac{G_{13}}{\lambda_1},
\een
which solves the equations for $G_{12},G_{13}$ and $G_{23}$ with $\sgn\left(\frac{G_{12}G_{13}}{G_{23}}\right)$ specifying the overall sign. The remaining three equations for $G_{01},G_{02}$ and $G_{03}$ are then solved by the two relations (\ref{gcondition}) and 
\ben
\lambda_0=\sgn\left(\frac{G_{12}G_{13}}{G_{23}}\right)\frac{G_{01}}{\lambda_1}.
\een
\begin{flushright}
$\Box$
\end{flushright}

The assumption $G_{ab}\neq 0$ is necessary: One solution to (\ref{gcondition}) is $G_{12}=G_{23}=G_{13}=0$ with the other $G_{ab}$ non-zero, which cannot be expressed as $G_{ab}=\pm\lambda_a\lambda_b$.

Further constraints, in addition to the linear simplicity constraints (\ref{zerorighthandside}), are needed to complete the identification $\Sigma^{AB}=\pm E^A \wedge E^B$. One possibility is to use the quadratic volume constraints (\ref{gcondition}). Take the three volume constraints (\ref{quadvol}),
\ben
\epsilon_{ABCD}\Sigma^{AB}_{01}\Sigma^{CD}_{23}=\epsilon_{ABCD}\Sigma^{AB}_{02}\Sigma^{CD}_{13}=\epsilon_{ABCD}\Sigma^{AB}_{01}\Sigma^{CD}_{23},
\een
and substitute the solution $\Sigma^{AB}_{ab}=G_{ab} e^{[A}_{a} e^{B]}_{b}$ of (\ref{zerorighthandside}). This gives precisely (\ref{gcondition}). The non-degeneracy assumption needed for (\ref{gcondition}) is then the usual one, namely $V\neq 0$ in (\ref{quadconst}).

This shows that imposing the linear version of the diagonal and off-diagonal simplicity constraints (\ref{zerorighthandside}) together with the quadratic volume constraints (\ref{quadvol}) and a non-degeneracy assumption on $\Sigma^{AB}$ implies that
\ben
\Sigma^{AB}=\pm c E^A \wedge E^B
\label{solucion}
\een
for some set of one-forms $E^A$, where $c>0$ can be chosen at will. Thus, linearising the diagonal and off-diagonal simplicity constraints means that two of the four types of solutions for $\Sigma^{AB}$ are removed, but on the other hand one needs to introduce a basis of three-forms $n_A$ at each spacetime point, which is put in as an additional variable. One also still has to assume $V\neq 0$. 

There is also generically no evolution of initial data with $V\neq 0$ into a degenerate $\Sigma^{AB}$ with $V=0$ and a non-geometric interpretation (this is part of the discussion of \cite{reisenclass}). The geometry of the spacetime manifold is specified by $E^A$ and not by $e^A$ which is only used to determine normals in the constraints.

Alternatively, one might prefer to use a linear version of the volume constraints as well. Consider the original equation (\ref{betterform})
\ben
n_{Adef} \Sigma^{AB}_{ab} = (\det e_a^A) \left(\epsilon_{adef}e_b^B-\epsilon_{bdef}e^B_a\right),
\een
which was equivalent to $\Sigma^{AB}=e^A\wedge e^B$ for an invertible frame field. So far we only considered one half of these equations, namely those with $\{a,b\}\subset\{d,e,f\}$. The other half have the form
\ben
n_{Abef} \Sigma^{AB}_{ab} = (\det e_a^A) \epsilon_{abef}e_b^B,\quad\mbox{no sum over }b,
\een
with $\epsilon_{abef}\neq 0$. One way to read these equations is as the requirement on the left-hand side to be totally antisymmetric in $(a,e,f)$:
\ben
n_{Abef} \Sigma^{AB}_{ab} = n_{Abfa} \Sigma^{AB}_{eb} = n_{Abae} \Sigma^{AB}_{fb}.
\label{linearvol}
\een

We could again try to turn the argument around and impose (\ref{linearvol}) as constraints on a $\frak{g}$-valued two-form $\Sigma^{AB}$ together with the linear simplicity constraints (\ref{zerorighthandside}). Substituting the solution $\Sigma^{AB}_{ab}=G_{ab}e^{[A}_a e^{B]}_b$ of the linear simplicity constraints into (\ref{linearvol}) gives (after diving by a non-zero factor $\half \det(e_a^A)$)
\ben
\epsilon_{abef}G_{ab}e_b^B = \epsilon_{ebfa}G_{eb}e_b^B = \epsilon_{fbae}G_{fb}e_b^B.
\label{multipleofe}
\een
For $\epsilon_{abef}\neq 0$ this would imply $G_{ab}=G_{eb}=G_{fb}$. By Claim 2, imposing (\ref{linearvol}) for one fixed $b$, say $b=0$, is generically not sufficient: If we know that $G_{01}=G_{02}=G_{03}\neq 0$, we still have the condition
\ben
G_{12}=G_{13}=G_{23},
\een
so that one would need more conditions of the form (\ref{linearvol}). These will then imply that all $G_{ab}$ are equal, $G_{ab}=c'$ for some $c'$ that could be positive, negative, or zero. One can absorb $|c'|$ by an overall redefinition, so that one has
\ben
\Sigma_{AB}=\pm c E^A\wedge E^B,
\een
for any chosen $c$, as before. Note that here it is possible, if $c'=0$ at a point, that all $E^A$ are zero this point and so $\Sigma^{AB}=0$ as well. While this is a very degenerate geometry, it is still a geometry. 

While the conditions (\ref{linearvol}), imposed for all values of $b$, are therefore sufficient to complete the identification of the two-form field $\Sigma^{AB}$ as $\pm c E^A\wedge E^B$, note that (\ref{linearvol}) is a massively redundant set of constraints: In order to obtain at most five relations on the coefficents $G_{ab}$ (two relations (\ref{gcondition}) if all $G_{ab}$ are nonzero), we are imposing {\em eight vector} equations! We have not exploited the fact that (\ref{multipleofe}) is a multiple of one of the vectors $e_b^B$, which are by assumption linearly independent. We could add several of the equations (\ref{linearvol}) for different $b$, instead of considering all equations for different $b$ separately. Let us try to impose
\ben
\sum_b \sum_{\{a,f\}\not\in\{b,e\}} n_{Abef} \Sigma^{AB}_{ab} = 0,\quad e\in\{0,1,2,3\}\mbox{ fixed}.
\label{sumconst}
\een
Again substituting the solution $\Sigma^{AB}_{ab}=G_{ab}e^{[A}_a e^{B]}_b$ of the linear simplicity constraints into (\ref{sumconst}), we obtain
\ben
\half\det(e_a^A) \sum_b \sum_{\{a,f\}\not\in\{b,e\}} \epsilon_{abef}G_{ab}e_b^B = 0,\quad e\mbox{ fixed},
\een
which implies, by linear independence of the $e_b^B$, that indeed $G_{ab}=G_{fb}$ for all $e\not\in\{a,b,f\}$. It is then sufficient to impose the constraint (\ref{sumconst}) for three different choices of $e$, say $e=0,1,2$, so that we only need three vector equations instead of eight. 

\

By absorbing the constant $G_{ab}=c'$ (all $G_{ab}$ are equal) we rescale all $e^A$ by the same factor to obtain the variables $E^A$ that will have the physical intepretation of frame fields encoding the metric geometry of spacetime. While in the case of quadratic volume constraints the one-forms $e^A$, or alternatively the three-forms $n^A$, only specified the normals to submanifolds $\{x^a=\const\}$, for linear volume constraints they can be directly interpreted, up to a position-dependent normalisation, as specifying an orthonormal basis in the cotangent space.

Note that this implies that one can assume a convenient normalisation for the one-forms $e^A$. Instead of just assuming non-degeneracy $\det(e^A_a)\neq 0$, one could fix $\det (e^A_a)=1$. This is no restriction of the physical content of the theory as the $e^A$, for both linear and quadratic volume constraints, only have a geometric interpretation after rescaling. One could then interpret $e^A_a$ as a map into $SL(4,\mathbb{R})$. For linear volume constraints, the relation between the normalised one-forms $e^A$ and the variables $E^A$ that are interpreted as frame fields is a single function on spacetime which may be viewed as a ``gauge" in the sense of Weyl \cite{spacetimematter}.

In contrast to the case of the quadratic volume constraint, no non-degeneracy assumption on the two-form $\Sigma^{AB}$ is needed to enforce simplicity. One might get $\Sigma^{AB}=0$ in some region as a solution to the constraints, in which case the action for this region will be zero. This is analogous to a metric with vanishing determinant in general relativity and, in contrast to the requirement $V\neq 0$ outlined above, not an additional issue. Notice, however, that one still has to assume that the tetrad field $e^A$ and, equivalently, the co-tetrad field $n_A$ are non-degenerate, in order for the simplicity and volume constraints to imply (\ref{solucion}). Failing this, one gets solutions of the constraints that admit no proper geometric interpretation.

\

In the end, writing the action for BF theory in terms of $\Sigma^{AB}$,
\ben
S = \int B^{AB}\wedge R_{AB} = \int \Sigma^{AB}\wedge R_{AB} + \frac{\gamma}{2}{\epsilon^{AB}}_{CD} \Sigma^{CD}\wedge R_{AB},
\een
we substitute (\ref{solucion}) into this action, which gives (setting $c=\frac{1}{8\pi\gamma G}$)
\ben
S = \frac{1}{8\pi G}\int_{\mathcal{R}\times\mathbb{R}}\sigma(x)\left(\half\epsilon_{ABCD}E^A\wedge E^B\wedge R^{CD}+\frac{1}{\gamma}E^A\wedge E^B\wedge R_{AB}\right).
\een

One is left with a field $\sigma(x)$ that can take the values $\pm 1$, but in the classical theory one may again argue that if $\sigma=1$ everywhere on an initial hypersurface, there will be no evolution into $\sigma=-1$. What we obtain is first order general relativity where one uses $(\det e)=\pm|\det e|$ instead of $|\det e|$ as a volume element in the action. If $\sigma$ is continuous as classical fields usually are assumed to be, this differs from the action with $|\det e|$ by an overall sign at most.

\

To summarise, we have identified both a linear version of the quadratic simplicity constraints and a linear version of the (quadratic) volume constraints in the continuum, which can be used to reduce topological BF theory to 4d gravity in the continuum. We have found also that both linear versions are slightly stronger (\ie more restrictive) than the corresponding quadratic constraints, so that the resulting constrained theory is likely to be closer to gravity at the quantum level than the one in which quadratic constraints are implemented. We now discuss the discrete counterpart of the constraints found above.

\subsection{Discrete Linear Constraints and their Relations}
The discrete analogue of (\ref{zerorighthandside}) is just the linear constraint used in \cite{eprlong,freidelkrasnov}, as desired:
\ben
n_A(\tetrahedron)\Sigma^{AB}(\triangle)=0\quad\forall\triangle\subset\tetrahedron.
\een

One could write down also a discrete version of (\ref{linearvol}), obtained in the natural way, demanding that within the same 4-simplex
\ben
n_{A}(\tetrahedron) \Sigma^{AB}(\triangle') = n_{A}(\tetrahedron') \Sigma^{AB}(\triangle'') 
\label{discvol}
\een
whenever $\triangle'\not\subset\tetrahedron$ and $\triangle''\not\subset\tetrahedron'$ and the edge shared by $\triangle'$ and $\tetrahedron$ is the same as that shared by $\tetrahedron'$ and $\triangle''$. 

In the following we adopt the notation of \cite{consistent}, where the tetrahedra in a given 4-simplex are labelled by ${\bf A,B,C,D,E}$, so that triangles are represented by ${\bf AB, AC}$, etc., and edges by combinations ${\bf ABC, ABD}$, etc. The orientation of the triangles and tetrahedra in (\ref{discvol}) is then fixed by the signs of the permutations of the letters,
\ben
n_{A}(\tetrahedron_{\bf A}) \Sigma^{AB}(\triangle_{\bf BC}) = -n_{A}(\tetrahedron_{\bf B}) \Sigma^{AB}(\triangle_{\bf AC}) = n_{A}(\tetrahedron_{\bf C}) \Sigma^{AB}(\triangle_{\bf AB}),\quad\mbox{etc.}
\label{linvol}
\een

In analogy to the continuum case, it will be sufficient to impose, instead of the full set of conditions (\ref{linvol}), certain linear combinations of (\ref{linvol}) to complete the geometric interpretation of the bivectors $\Sigma^{AB}(\triangle)$. The discrete analogue of the three continuum equations (\ref{sumconst}), where the index $e$ was kept fixed, is to pick one of the tetrahedra and add those six equations out of (\ref{linvol}) which involve triangles belonging to this tetrahedron. Starting with ${\bf A}$, we impose the constraint
\ben
\sum_{\{i,j\}\not\ni A}n_{A}(\tetrahedron_{\bf i})\Sigma^{AB}(\triangle_{\bf Aj})=0,
\label{discsumc}
\een
and the equivalent conditions for the tetrahedra ${\bf B}$ to ${\bf E}$, thereby needing to satisfy only five instead of 20 volume constraints.

The above discrete formulation of the linearised volume constraints resembles strongly the edge simplicity constraints studied, in a canonical setting, in \cite{biancajimmy}, and it imposes indeed the same restriction on the discrete data. However, it does not match exactly any of the various expressions given for these edge simplicity constraints in \cite{biancajimmy}. The correspondence between the two, therefore, deserves to be studied in more detail, given also that edge simplicity constraints have been shown to be crucial for the kinematical phase space of BF theory (and of loop gravity) to reduce to that of discrete gravity, in accordance with what we find here in a covariant setting.

\

In spin foam models such as \cite{eprlong,freidelkrasnov}, as we mentioned earlier, only the diagonal and off-diagonal simplicity constraints, but no quadratic volume constraints (\ref{quadvol}) are imposed. This is because in the discrete setting, one can use the closure constraint (\ref{closure}), imposed in all the tetrahedra in a 4-simplex, to relate the (quadratic) simplicity constraints to the volume constraints, so that if the former are imposed everywhere the latter follows. Since the quadratic simplicity constraints follow from the linear ones, as can be easily checked, this argument is still valid if one uses linear simplicity constraints.

One might hope that the sufficient set of linear volume constraints (\ref{discsumc}) would also follow from the linear simplicity constraints and the closure constraints. This is almost the case, but not quite. In fact, one more constraint should be added to simplicity and closure imposed in the five tetrahedra in the 4-simplex. This is a ``4d closure'' constraint of the form
\ben
n_A(\tetrahedron_{\bf A})+n_A(\tetrahedron_{\bf B})+n_A(\tetrahedron_{\bf C})+n_A(\tetrahedron_{\bf D})+n_A(\tetrahedron_{\bf E})=0,
\label{nclosure}
\een
where $\tetrahedron_i$ are the (appropriately oriented) tetrahedra of a given 4-simplex. 

Just as for the usual closure constraint (\ref{closure}), there are two ways to understand why such a constraint must be imposed. Recall that if one demands the triangles described by discrete variables $\Sigma^{AB}_{\triangle}$ close to form a tetrahedron, they have to satisfy (\ref{closure}). Alternatively, one can start with the continuum field equation $\nabla_{[a}^{(\omega)}\Sigma^{AB}_{bc]}=0$, where $\nabla^{(\omega)}$ is the covariant derivative for the connection $\omega^{AB}$, set the (flat) connection to zero by a gauge transformation, and integrate this over an infinitesimal 3-ball (whose triangulation is a tetrahedron).

The new constraint (\ref{nclosure}) seems to be the analogous statement that tetrahedra close up to form a 4-simplex. By Hodge duality $\wedge^1\mathbb{R}^4\simeq\wedge^3\mathbb{R}^4$ and any internal covector $n_A(\tetrahedron)$ can be mapped to a three-form; unlike for two-forms, any three-form can be written as $e^1\wedge e^2\wedge e^3$ for some $e^\alpha$. Demanding that the tetrahedra described by these three-forms form a closed surface is then (\ref{nclosure}). Thus the simplicial geometric reasoning goes through also for this new constraint. In terms of the equations of motion of the theory, on the other hand, the only argument for the need of this constraint is the following. If $\nabla_{[a}^{(\omega)}\Sigma^{AB}_{bc]}=0$ and we assume that $\Sigma^{AB}=\pm e^A\wedge e^B$, then it follows that $\nabla_{[a}^{(\omega)}n^A_{bcd]}=0$ as well. Integrating this equation (with the connection again set to zero) over a 4-ball (which can be thought of as our 4-simplex) whose boundary is a 3-sphere, triangulated by tetrahedra, then leads to (\ref{nclosure}). We then however have to assume simplicity of $\Sigma^{AB}$. A more direct derivation of (\ref{nclosure}) from the equations of motion would be desirable.

The role of this constraint, anyway, is the following. Consider the closure constraint
\ben
\Sigma^{AB}(\triangle_{\bf AB})+\Sigma^{AB}(\triangle_{\bf AC})+\Sigma^{AB}(\triangle_{\bf AD})+\Sigma^{AB}(\triangle_{\bf AE})=0.
\label{aclosure}
\een
Contracting with $n_A(\tetrahedron_{\bf B})$ gives, using the linear simplicity constraint $n_A(\tetrahedron_{\bf B})\Sigma^{AB}(\triangle_{\bf AB})=0$,
\ben
n_A(\tetrahedron_{\bf B})\Sigma^{AB}(\triangle_{\bf AC})+n_A(\tetrahedron_{\bf B})\Sigma^{AB}(\triangle_{\bf AD})+n_A(\tetrahedron_{\bf B})\Sigma^{AB}(\triangle_{\bf AE})=0.
\label{clocon1}
\een
Alternatively, one may start with the 4d closure constraint and contract with $\Sigma^{AB}(\triangle_{\bf AB})$ to get, again using the linear simplicity constraints,
\ben
n_A(\tetrahedron_{\bf C})\Sigma^{AB}(\triangle_{\bf AB})+n_A(\tetrahedron_{\bf D})\Sigma^{AB}(\triangle_{\bf AB})+n_A(\tetrahedron_{\bf E})\Sigma^{AB}(\triangle_{\bf AB})=0.
\label{clocon2}
\een
In total one obtains 20 + 10 = 30 equations of this kind that can be used to express some of the combinations $n_A(\tetrahedron)\Sigma^{AB}(\triangle)$ in terms of others. Substituting the resulting expressions into the five discrete volume constraints (\ref{discsumc}) one finds that the equations (\ref{discsumc}) indeed follow from the relations (\ref{clocon1}) and (\ref{clocon2}). We have seen in the continuum that the summed constraints (\ref{sumconst}) are sufficient to identify $\Sigma^{AB}=\pm E^A\wedge E^B$, and hence we find that in the discrete case the situation is analogous to the case of quadratic constraints in that a sufficient set of volume constraints can be viewed as secondary.

To see more clearly what happens in both our construction and in the case of quadratic constraints analyzed in \cite{consistent}, note that in our linear case one could use the 3d and 4d closure constraints to express the variables $n_A(\tetrahedron_{\bf E})$ and $\Sigma^{AB}(\triangle_{\bf AE}), \Sigma^{AB}(\triangle_{\bf BE}), \Sigma^{AB}(\triangle_{\bf CE}), \Sigma^{AB}(\triangle_{\bf DE})$ in terms of the others. Taking the linear simplicity constraints into account, one is then left with twelve independent combinations $n_A(\tetrahedron)\Sigma^{AB}(\triangle)$, just as in the continuum. In the continuum, we saw that one can impose the three additional constraints (\ref{sumconst}) on the twelve contractions $n_{Abef}\Sigma^{AB}_{ab}$ to complete the identification $\Sigma^{AB}=\pm E^A\wedge E^B$. In the discrete case, one has the following three additional conditions coming from linear cross-simplicity constraints:
\bea
0=n_A(\tetrahedron_{\bf E})\Sigma^{AB}(\triangle_{\bf AE})&=&n_A(\tetrahedron_{\bf B})\Sigma^{AB}(\triangle_{\bf AC})+n_A(\tetrahedron_{\bf B})\Sigma^{AB}(\triangle_{\bf AD})+n_A(\tetrahedron_{\bf C})\Sigma^{AB}(\triangle_{\bf AB})\nn
\\& & + n_A(\tetrahedron_{\bf C})\Sigma^{AB}(\triangle_{\bf AB})+n_A(\tetrahedron_{\bf D})\Sigma^{AB}(\triangle_{\bf AB})+n_A(\tetrahedron_{\bf D})\Sigma^{AB}(\triangle_{\bf AC})
\eea
and similar ones coming from $n_A(\tetrahedron_{\bf E})\Sigma^{AB}(\triangle_{\bf BE})=0$ and $n_A(\tetrahedron_{\bf E})\Sigma^{AB}(\triangle_{\bf CE})=0$. These are precisely the analogue of the continuum constraints (\ref{sumconst}).

Similarly, in the case of quadratic simplicity constraints, one can use 3d closure to eliminate $\Sigma^{AB}(\triangle_{\bf AE}), \Sigma^{AB}(\triangle_{\bf BE}), \Sigma^{AB}(\triangle_{\bf CE}), \Sigma^{AB}(\triangle_{\bf DE})$. Then one observes that additional quadratic cross-simplicity constraints give expressions such as
\bea
0 = \epsilon_{ABCD}\Sigma^{AB}(\triangle_{\bf AE})\Sigma^{CD}(\triangle_{\bf BE})=\epsilon_{ABCD}\Sigma^{AB}(\triangle_{\bf AC})\Sigma^{CD}(\triangle_{\bf BD})+\epsilon_{ABCD}\Sigma^{AB}(\triangle_{\bf AD})\Sigma^{CD}(\triangle_{\bf BC})
\eea
which are equivalent to the desired (two) volume constraints.  

All of this is an exercise in solving a system of linear equations for which there might be a more simple and elegant description, but the upshot is the following. The sufficient set of linear volume constraints (\ref{discsumc}) does indeed follow from the linear simplicity constraints and the closure constraints, once one also imposes a four-dimensional closure constraint on the normals to tetrahedra that seems very natural in light of their geometric interpretation. Just as in the formulation in terms of quadratic simplicity constraints \cite{consistent}, the volume constraints can be viewed as secondary constraints that imply conservation of the simplicity constraints in time, or put differently, the volume constraints follow if the simplicity constraints hold everywhere. Once more this strenghtens the relationship between the discrete linear volume constraints we have identified and the edge simplicity constraints of \cite{biancajimmy}.

\sect{Lagrangian and Hamiltonian Formulation}
Let us briefly outline the Lagrangian formulation of 4d gravity resulting from our linear constraints added to BF theory.
One adds the linear simplicity and volume constraints to the action of BF theory using Lagrange multipliers:
\ben
S = \int d^4 x\;\left(\frac{1}{4}\epsilon^{abcd}\Sigma^{AB}_{ab}R_{ABcd}[\omega]+\frac{\gamma}{8}\epsilon^{abcd}\epsilon_{ABCD}\Sigma^{AB}_{ab}R^{CD}_{cd}[\omega]+\Xi_B^{abdef}n_{Adef}\Sigma^{AB}_{ab}\right),
\label{newakshn}
\een
where the Lagrange multiplier field $\Xi_B^{abdef}$ satisfies $\Xi_B^{abdef}\equiv \Xi_B^{[ab][def]}$, and
\ben
\epsilon_{aef}\Xi_B^{abbef}=0\quad\mbox{(no sum over }b).
\een
Indeed, varying with respect to $\Xi_B$ then gives back the constraints
\ben
n_{Adef}\Sigma^{AB}_{ab} = \cases{0, & $\{a,b\}\subset\{d,e,f\}$, \cr \epsilon_{aef}f_b^B, & $b=d$ (for some $f_b^B$).}
\een
Note that the second line corresponds to the set of constraints (\ref{linearvol}) and not to the summed version (\ref{sumconst}), and that it is clearly sufficient for the geometric interpretation of $\Sigma^{AB}$. The field equation from varying with respect to the connection $\omega$ is the usual
\ben
\nabla^{(\omega)}_{[a} \Sigma^{AB}_{bc]}=0,
\een
where $\nabla$ is the covariant derivative for the connection $\omega^{AB}$. The remaining equations involve the Lagrange multipliers, as would be expected:
\ben
\frac{1}{4}\epsilon^{abcd}R_{ABcd}[\omega]+\frac{\gamma}{8}\epsilon^{abcd}\epsilon_{ABCD}R^{CD}_{cd}[\omega]+\Xi_{[B}^{abdef}n_{A]def}=0,\qquad \Xi_B^{abdef}\Sigma^{AB}_{ab}=0.
\een
We have seen that the constraints imply that $\Sigma^{AB}=\pm E^A\wedge E^B$, and when substituting this back into the action one will recover general relativity, modulo the possible sign ambiguity we have already discussed.

\

We leave a complete Hamiltonian analysis of this theory to future work. However, we note a feature of the theory that follows directly from the use of linear constraints, and from the introduction of the additional variables $n_A$. 

As in unconstrained BF theory the initial dynamical variables will be the spatial part of the connection $\omega^{AB}_k$ and its conjugate momentum $P_{AB}^k\equiv\frac{1}{2}\epsilon^{ijk}\Sigma_{ABij}$. We also saw that the equation of motion $\nabla^{(\omega)}_{[a} \Sigma^{AB}_{bc]}=0$ is unaffected by the constraints. Hence there will be Gauss constraints of the form
\ben
\mathcal{G}^{CD}\equiv\partial_i P^{CDi}+{\omega^C}_{Ei}P^{EDi}+{\omega^D}_{Ei}P^{CEi}
\label{gauss}
\een
on the canonical momenta. Their role is to generate $G$ gauge transformations. 

Looking at the action (\ref{newakshn}), one would already require that (\ref{gauss}) should be modified to generate gauge transformations on the normals $n_{Adef}$; (\ref{newakshn}) is only invariant under gauge transformations if the three-forms $n_{Adef}$ are transformed. The need for such a modification is also seen if one computes Poisson brackets between the linear simplicity and Gauss constraints. Define ``smeared" constraints
\ben
C[\Xi]:=\int\Xi_B^{ij,def}n_{Adef}\epsilon_{ijk}P^{ABk},\quad\mathcal{G}[\Lambda]:=\int\Lambda^{CD}\mathcal{G}_{CD}.
\een
One then finds that
\bea
\{C[\Xi],\mathcal{G}[\Lambda]\} & = & -\int\frac{\delta C[\Xi]}{\delta P^{GHm}}\frac{\delta \mathcal{G}[\Lambda]}{\delta \omega_{GHm}}\nn
\\& = & - \int \Xi_B^{ij,def}n_{Adef}\epsilon_{ijm}\left[\Lambda^{AD}{{P^B}_D}^m-\Lambda^{BD}{{P^A}_D}^m\right]\nn
\\& = & -C[\Lambda\cdot\Xi]-\int \Xi_B^{ij,def}\Lambda^{AD}n_{Adef}\epsilon_{ijm}{{P^B}_D}^m,
\eea
where $(\Lambda\cdot\Xi)_D^{ij,def}={\Lambda_D}^B \Xi_B^{ij,def}$. The first term alone would imply that $\mathcal{G}[\Lambda]$ generates gauge transformations, but the second term is an unwanted extra piece. For $\mathcal{G}$ to be a generator of gauge transformations, it must be first class (\ie commute with other constraints up to linear combinations of constraints). We can remedy this by adding the variables $n_{Aabc}$ to the phase space, together with their conjugate momenta $\pi^{Aabc}$. This extension of the phase space is analogous to the situation considered in \cite[sect. 3]{sergeikirill} for a generalised quadratic constraint formulation of Plebanski theory. Now we can define a new Gauss constraint
\ben
\mathcal{G}'^{CD}\equiv\mathcal{G}^{CD}-n^{[C}_{abc}\pi^{D]abc}.
\een
Then, computing the Poisson brackets of the new Gauss constraint with $C[\Xi]$, one finds
\ben
\{C[\Xi],\mathcal{G}'[\Lambda]\} = \{C[\Xi],\mathcal{G}[\Lambda]\} -\int \Xi_B^{ij,def}\Lambda^{CA}n_{Cdef}\epsilon_{ijk}P^{ABk}=-C[\Lambda\cdot\Xi],
\een
as desired. We have however increased the number of phase space variables at each point by 32.

\

A similar reformulation of the Gauss constraint, leading to a relaxation of the gauge invariance properties of spin network states, has been already suggested by the Hamiltonian analysis of the Plebanski theory \cite{henneaux}, and it has been advocated in the loop quantum gravity context in \cite{projected, sergeietera} as well as spin foam and group field theory context \cite{sergei, danieleAristide, danieleGFtsimpl}.

\sect{Summary and Outlook}
We have investigated a formulation of classical BF-Plebanski theory where the constraint $\Sigma^{AB}=\pm e^A\wedge e^B$, needed to reproduce general relativity in four dimensions, starting from topological BF theory, is imposed through constraints linear in the bivector field $\Sigma^{AB}$. 
The discrete counterpart of a part of these linear constraints (the \lq simplicity constraints\rq), in fact, has proven very useful in the spin foam approach to quantum gravity \cite{eprlong,freidelkrasnov, eprl}. 

\

The corresponding continuum constraints have been easily identified, and can indeed be used to replace the quadratic ``diagonal'' and ``off-diagonal'' parts of the simplicity constraints appearing in the Plebanski formulation. As in the discrete case, one needed to introduce a new set of variables $n_A$ which are assumed to form a basis of three-forms at each point of spacetime, and are slightly stronger than the quadratic constraints: they eliminate two of the four sectors of solutions that are present for quadratic constraints.

In the second part of the analysis we found that the quadratic volume constraints of the Plebanski formulation, needed to complete the identification $\Sigma^{AB}=\pm e^A\wedge e^B$, can also be replaced by linear constraints, which again are stronger than their quadratic analogues. They do not require an additional non-degeneracy assumption on $\Sigma^{AB}$. However, a non-degeneracy assumption on the three-forms $n_A$ is still necessary, and only when this is imposed one can hope to eliminate all ``non-geometric'' degenerate configurations for $\Sigma^{AB}$, which are feared to dominate the quantum theory in the case of quadratic volume constraints. Also, while for quadratic volume constraints the variables $n_A$ merely specify normals to submanifolds $\{x^a=\const\}$ and hence can be independently rescaled arbitrarily at each point, for linear volume constraints they directly specify, up to an overall rescaling, the frame field encoding the metric geometry, \ie an orthonormal basis in the cotangent space at each spacetime point.

We have then analyzed the discrete (simplicial) translation of the linear constraints we identified. In the context of spin foams, the quadratic volume constraints follow from imposing the (quadratic) diagonal and off-diagonal simplicity constraints everywhere together with closure constraints on the discrete variables $\Sigma^{AB}(\triangle)$. We have shown a similar property for the linear volume constraints. If (linear) diagonal and off-diagonal simplicity constraints and closure constraints for {\it both} bivector variables $\Sigma^{AB}(\triangle)$  {\it and} normals $n_A(\tetrahedron)$ are imposed everywhere, a sufficient set of linear volume constraints follows. This means that ``non-geometric'' bivector configurations cannot appear if the additional closure constraint on the normals holds, and the same normals are assumed to be non-degenerate.

\

We have not performed a complete Hamiltonian analysis of the resulting linear constrained BF action for gravity, but only noted that the use of linear simplicity and volume constraints immediately requires a modification of the usual Gauss constraint to generate a transformation of the normal 3-form variables $n_A$ alongside that of the $\Sigma$'s; a similar relaxation of the Gauss constraint, which translates at the spin foam and discrete gravity level into a closure constraint for simplices, and in the canonical quantum gravity context into a generalisation of spin network states, has been suggested on more than one occasion in the literature \cite{projected, sergeietera, sergei, danieleAristide}, even if its proper implementation at the quantum level has not been yet developed. On the classical level, therefore, a full Hamiltonian analysis of the constraints would be highly desirable. This would involve adding momenta for the components $\Sigma^{AB}_{0i}$, which are Lagrange multipliers in unconstrained BF theory, as well as those for the normals $n_{A}$ we have introduced, so that all variables can transform nontrivially under $G$ gauge transformations generated by a modified Gauss constraint, as shown.

Still at the classical level, but with obvious implications for the quantisation, one aspect of our construction that deserves further work is the relation between the discretised linear volume constraints we have found and the edge simplicity constraints used in \cite{biancajimmy}, in turn related to the conditions on connection variables of \cite{valentin}. As noted, the two sets of constraints appear to be very similar, and their role in the classical theory is the same, in particular, they remove (partly) the non-geometric configurations from the configuration space (or phase space) of the theory and appear as ``secondary'' in the sense specified above. So it natural to conjecture that one is simply a reformulation of the other. The implications for the quantum theory are not only due to the dominant role that non-geometric configurations may play in the quantum theory, if not removed, but also in the fact that  one discrete formulation of these constraints can actually be simpler to implement in a spin foam context than the other.

The possible use of our findings in the spin foam and group field theory context, and more generally in any quantisation based on the formulation of gravity as a constrained BF theory, are in fact most interesting. In particular, it seems to be important to explore how a closure constraint on normals could be implemented into existing spin foam models, given that we found it to be necessary for the full imposition of the geometric constraints on the variables of topological BF.  A convenient setting to do so could be the GFT formulation of \cite{danieleAristide}, since there the simplicial geometry and the contact with classical actions is brought to the forefront.

\section*{Acknowledgments}
SG was supported by EPSRC and Trinity College, Cambridge, partly through a Rouse Ball Travelling Studentship in Mathematics. DO gratefully acknowledges financial support from the A. von Humboldt Stiftung, through a Sofja Kovalevskaja Prize. We thank S. Alexandrov, A. Baratin, B. Dittrich and J. Ryan for discussions and comments on this work.

\end{document}